\documentclass[11pt,onecolumn,english,showpacs,preprintnumbers,amsmath,amssymb,floatfix]{revtex4-1}

\usepackage{lineno}

\usepackage[T1]{fontenc}
\usepackage[latin9]{inputenc}
\usepackage{color}
\usepackage{array}
\usepackage{amstext}
\usepackage{graphicx}
\usepackage{esint}
\usepackage{rotating}
\usepackage[font=small,labelfont=bf]{caption}
\usepackage[colorlinks = true,
linkcolor = blue,
urlcolor  = magenta,
citecolor = red,
anchorcolor = blue]{hyperref}

\makeatletter

\@ifundefined{textcolor}{}
{%
 \definecolor{BLACK}{gray}{0}
 \definecolor{WHITE}{gray}{1}
 \definecolor{RED}{rgb}{1,0,0}
 \definecolor{GREEN}{rgb}{0,1,0}
 \definecolor{BLUE}{rgb}{0,0,1}
 \definecolor{CYAN}{cmyk}{1,0,0,0}
 \definecolor{MAGENTA}{cmyk}{0,1,0,0}
 \definecolor{YELLOW}{cmyk}{0,0,1,0}
 }

\@ifundefined{definecolor}
 {\usepackage{color}}{}
\@ifundefined{definecolor}
 {\usepackage{color}}{}
\makeatother
\usepackage{babel}

\newcommand{\madgraph}{{\sc MadGraph}}

\newcommand{\feynrules}{{\sc Feyn\-Rules}}
\newcommand{\pythia}{{\sc Pythia}}

\def\Re{{\cal R \mskip-4mu \lower.1ex \hbox{\it e}\,}}
\def\Im{{\cal I \mskip-5mu \lower.1ex \hbox{\it m}\,}}

\def\tev{\,{\ifmmode\mathrm {TeV}\else TeV\fi}}
\def\gev{\,{\ifmmode\mathrm {GeV}\else GeV\fi}}
\def\mev{\,{\ifmmode\mathrm {MeV}\else MeV\fi}}
\def\to{\rightarrow}

\begin{document}

% \linenumbers

%%%%%%%%%%%%%%%%%%%%%%%%%%%%%%

%%  Titles  %% 

\title { Single top quark production as a probe of anomalous $tq\gamma$ and $tqZ$ couplings at the FCC-ee  }

\author { Hamzeh Khanpour$^{1,2}$ }
\email{  Hamzeh.Khanpour@mail.ipm.ir  }

\author { Sara Khatibi$^{2}$ }
\email{ S.Khatibi@ipm.ir }

\author { Morteza Khatiri Yanehsari$^{3}$ }
\email{ Khatiri@ipm.ir  }

\author { Mojtaba Mohammadi Najafabadi$^2$ }
\email{ Mojtaba@ipm.ir  }

\affiliation {
$^{(1)}$Department of Physics, University of Science and Technology of Mazandaran, P.O.Box 48518-78195, Behshahr, Iran           \\ 
$^{(2)}$School of Particles and Accelerators, Institute for Research in Fundamental Sciences (IPM), P.O.Box 19395-5531, Tehran, Iran       \\
$^{(3)}$School of Physics, Institute for Research in Fundamental Sciences (IPM), P.O.Box 19395-5531, Tehran, Iran    }

\date{\today}

%
%%%%%%%%%%%%%%%%%%%%%%%%%%%%%%%%%%%%%%%%%%%%%%%%%%%%%%%%%%%
%
\begin{abstract}\label{abstract}

In this paper, a detailed study to probe the top quark Flavour-Changing Neutral Currents (FCNC) $tq\gamma$ 
and $tqZ$  at the future $e^{-}e^{+}$ collider FCC-ee in three different center-of-mass energies
of 240, 350 and 500 GeV is presented.
A set of useful variables are proposed and used in a multivariate technique to separate signal
$e^- e^+ \rightarrow Z/\gamma \rightarrow t \bar{q} ~ ( \bar{t} q )$ from standard model background processes. 
The study includes a fast detector simulation based on the {\sc delphes} package to consider the detector effects.
The $3 \sigma$ discovery regions and the upper limits on the FCNC branching ratios at 95\% confidence level (CL) in terms of the integrated luminosity are presented.
It is shown that with 300 fb$^{-1}$ of integrated luminosity of data,
FCC-ee would be able to exclude the effective coupling strengths above ${\cal O} (10^{-4}-10^{-5})$ which is corresponding to branching
fraction of ${\cal O}(0.01-0.001)$\%. We show that moving to a high-luminosity regime leads to a significant
improvement on the upper bounds on the top quark FCNC couplings to a photon or a $Z$ boson.

\end{abstract}


\maketitle

\tableofcontents{}

%
%%%%%%%%%%%%%%%%%%%%%%%%%%%%%%%%%%%%%%%%%%    Introduction    %%%%%%%%%%%%%%%%%%%%%%%%%%%%%%%%%%%%%%%%%%%%%%%%%
%
\section{Introduction}\label{Introduction}

The top quark with its large mass and very short life time is one of the most interesting
discovered particles in the Standard Model (SM). Studying the top quark enables us to
investigate the electroweak symmetry breaking mechanism (EWSB) as well as searching
for extensions of the SM.
In the framework of the SM, top-quark Flavour-Changing Neutral Currents (FCNC) only arise at loop level and are highly suppressed
because of the GIM (Glashow-Iliopoulos-Maiani) mechanism~\cite{Glashow:1970gm}.
For instance, the SM predictions for the branching fractions of FCNC processes like $t \rightarrow \gamma u (c)$ and $t \rightarrow Z u (c)$
are of the order of 10$^{-16} (10^{-14})$ and 10$^{-17} (10^{-14})$, respectively~\cite{Agashe:2013hma}. The ability of the present experiments is far from
measuring such tiny branching ratios. On the other hand, several extensions of the SM such as Technicolor, SUSY models, Higgs doublet models predict
much higher branching ratios up to $10^{8}-10^{10}$ order of magnitude larger than SM predictions~
\cite{Agashe:2013hma,Bejar:2008ub,Cao:2008vk,GonzalezSprinberg:2006am,Diaz:2001vj,Lu:2003yr,Couture:1997ep}.
Consequently, any observation of these rare FCNC transitions would be a clear signal of new physics beyond the SM.

So far, there are several experimental studies in searching for FCNC transitions of the top quark to a photon or a Z boson through different channels~\cite{CMS:2014hwa,Aad:2015uza,Aad:2015gea,Khachatryan:2015att,Guo:2016kea,TheATLAScollaboration:2013vha,Aad:2014dya,Chatrchyan:2012hqa,Aad:2012ij,Chatrchyan:2013nwa,CHAO:2014dqa,Abazov:2011qf,Abramowicz:2011tv,Obraztsov:1997if,Achard:2002vv,Abdallah:2003wf}.
The most stringent observed upper limits at 95\% confidence level (CL) have been found to be~\cite{CMS:2016bss,Aad:2015uza,Khachatryan:2015att}:
%-------------------------------------
\begin{eqnarray}
	{\rm CMS:}  && Br (t\rightarrow Zu)<0.017\%  \,, \nonumber   \\ 
	&& Br (t\rightarrow Zc)<0.020\% \,,   \nonumber   \\ 
	{\rm ATLAS:} && Br (t\rightarrow Zq) < 0.07\% ~ ({\rm observed}) \,,\nonumber  \\ 
    && Br (t\rightarrow Zq) < 0.08\% ~ ({\rm expected}) \,, \nonumber  \\ 
	{\rm CMS:}	&& Br (t \rightarrow u\gamma) < 0.013\% \,, \nonumber  \\ 
     && Br (t \rightarrow c\gamma) < 0.170\% \,. 
\end{eqnarray}
%-------------------------------------
It is notable that even at the future upgrades of the LHC,
these bounds would not be improved considerably. For example, the future upper bounds
on $Br (t \rightarrow qZ)$ have been predicted to be $0.01\%$ at $95\%$ CL at 14 TeV center-of-mass
energy with 3000 fb$^{-1}$ of integrated luminosity of
data~\cite{CMS:2013zfa,ATLAS-Collaboration:2012jwa}. The branching
fraction of $Br (t \rightarrow q\gamma)$ would be reachable down to
$10^{-4}$ for $q=c$ and $10^{-5}$ for $q=u$ at the integrated
luminosity of 3000 fb$^{-1}$ at the LHC~\cite{CMS-HL}.
Therefore, an important task is to look at the future colliders potential
to search for the anomalous FCNC couplings, in particular the $e^- e^+$ colliders such as International Linear Collider (ILC)~\cite{Fujii:2015jha,Behnke:2013lya,Barklow:2015tja,Asner:2013psa,Moortgat-Picka:2015yla,Asner:2013hla,Brau:2012hv,Martinez:2002st}, Compact Linear Collider  (CLIC)~\cite{Linssen:2012hp,Aicheler:2012bya,Abramowicz:2013tzc,Lebrun:2012hj},  Circular Electron-Positron Collider (CEPC)~\cite{CEPC-SPPCStudyGroup:2015csa,CEPC-SPPCStudyGroup:2015esa} and the  high-luminosity high-precision Future Circular Collider (FCC-ee)~~\cite{Gomez-Ceballos:2013zzn,Koratzinos:2015ywz,d'Enterria:2016yqx,Ellis:2015sca,Janot:2015yza,d'Enterria:2015toz,Benedikt:2015iia,Koratzinos:2015fya,Zimmermann:2015mea,d'Enterria:2016cpw}.

In Refs.~\cite{AguilarSaavedra:2001ab,AguilarSaavedra:2000db}, an analysis has been performed to probe the sensitivity of a future $e^- e^+$
collider to top quark FCNC to the photon and a Z boson in the $ e^- e^+ \rightarrow Z/\gamma \rightarrow t \overline{q} ~ ( \overline{t} q )$ channel.
This analysis has been done at the center-of-mass energies of 500 GeV and 800 GeV with the integrated luminosity of up to 1 ab$^{-1}$
without including the effects of parton showering, hadronization, and decay of unstable particles. However, the analysis
considers cases with and without the beam polarization to estimate the sensitivity to $tq \gamma$ and $tqZ$ FCNC couplings.

The future large scale circular electron-positron collider (FCC-ee) would be one of
the high-precision and high-luminosity machines which will be able to perform precise measurements on
the Higgs boson, top-quark, $Z$ and $W$ bosons~\cite{Gomez-Ceballos:2013zzn,Koratzinos:2013ncw}.
Due to the expected large amount of data and large production rates, FCC-ee
can provide an excellent opportunity for precise studies, in particular in the top quark sector.
FCC-ee is designed to be working at the center-of-mass energy up to
the $t \bar{t}$ threshold mass, i.e. $\sqrt{s} = 350$ GeV which is upgradeable to 500 GeV. The goal is to reach to a luminosity
of $L = 1.3 \times 10 ^{34}$ cm$^{-2}$s$^{-1}$~\cite{Gomez-Ceballos:2013zzn,Koratzinos:2013ncw}.

In this paper, our aim is to study the anomalous FCNC of $t q \gamma$ and $tqZ$ via single top quark
production in the FCC-ee at three different center-of-mass energies of 240 GeV, 350 GeV and 500 GeV.
The final state consists of a top quark in association with a light-quark.
We consider the leptonic decay of the $W$ boson in top quark decay, ($t\rightarrow Wb \rightarrow \ell \nu_\ell b$, where $\ell = e, \mu$).
In the analysis, we perform parton shower, hadronization and decays of unstable particles as well as the detector effects.
We present the $3 \sigma$ discovery ranges and upper limits on the branching ratios at 95\% C.L in terms of the integrated luminosity.
Finally, the results are compared with the present and future results from the LHC experiments.

The paper is organized as follows. In Section~\ref{Theoretical-formalism}, we present the theoretical framework
which describes the top quark FCNC couplings to a photon and a $Z$ boson.
The Monte Carlo event generation, detector simulation and signal separation from backgrounds are described in
Section~\ref{Monte-Carlo}. In Section~\ref{results}, the results of the sensitivity estimation are presented. Finally, Section~\ref{Conclusions} concludes the paper.

%
%%%%%%%%%%%%%%%%%%%%%%%%%%%%%%%%%%%%%%%%%%    Theoretical formalism    %%%%%%%%%%%%%%%%%%%%%%%%%%%%%%%%%%%%%%%%%%%%%%%%%
%
\section{ Theoretical formalism }\label{Theoretical-formalism}

\begin{figure}[tbh]
	\centerline{\includegraphics[width=0.50\textwidth]{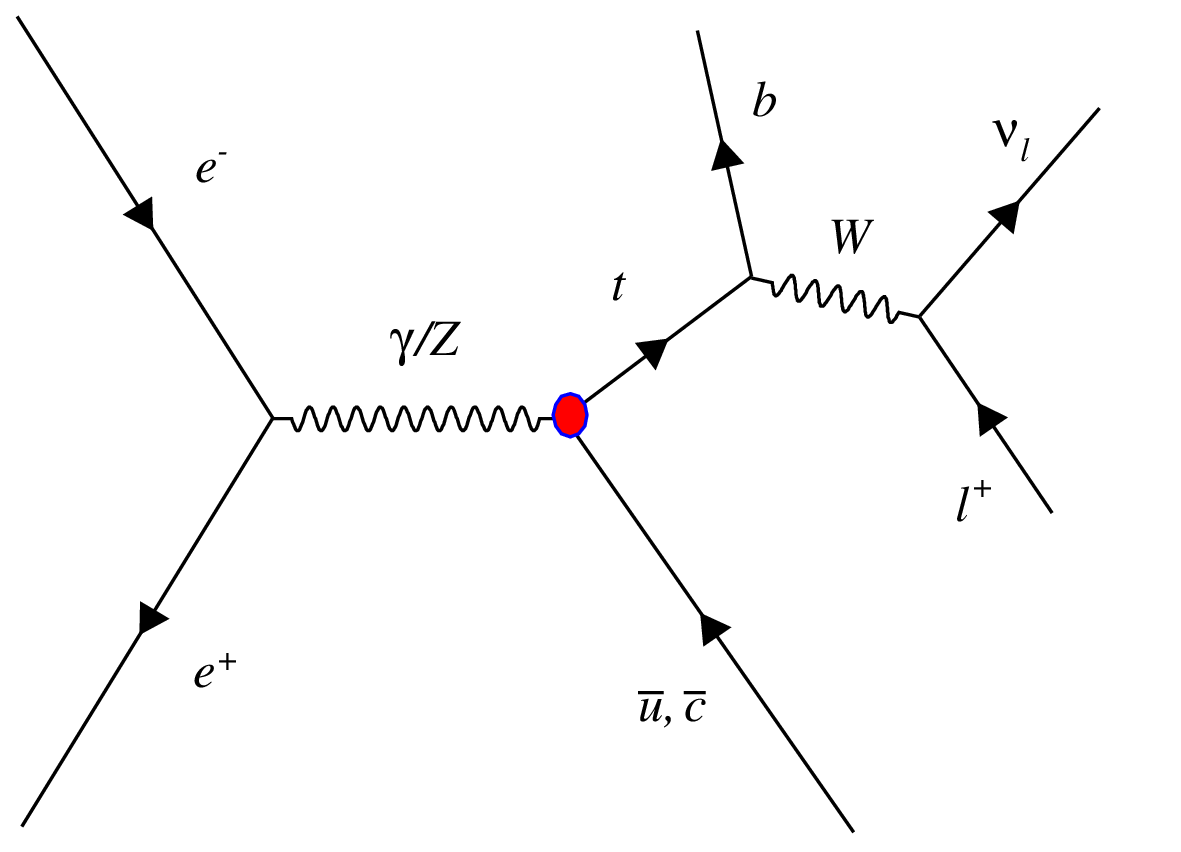}}
	\caption{The Feynman diagram for production of a top in association with a light quark due to the anomalous couplings $tq\gamma$
		and $tqZ$ in electron-positron collisions.}
	\label{feynman-diagram}
\end{figure}

The anomalous FCNC couplings of a top quark with a photon and a $Z$ boson can be written in a model independent way using an
effective Lagrangian approach. The lowest order terms describing $tq \gamma$ and $tqZ$ couplings have the following form
~\cite{Obraztsov:1997if,Malkawi:1995dm,Hosch:1997gz,AguilarSaavedra:2000db,AguilarSaavedra:2004wm}:
%--------------------------------
\begin{equation}\label{Effective-Lagrangian}
	\begin{split}
		\mathcal{L}_{\rm eff} = & \ \sum_{q = u, c} \bigg[
		e \lambda_{tq} \bar{t} (\lambda^{v}-\lambda^{a}\gamma^{5})\frac{i \sigma _{\mu \nu }  q^{\nu}}{m_{t}} qA^{\mu}  \\
		&\ + \frac{g_W}{2 c_W}  \kappa_{tq} \bar{t} (\kappa^{v}-\kappa^{a}\gamma^{5})\frac{ i\sigma_{\mu \nu} q^{\nu}}{m_{t}} q\ Z^{\mu}  \\
		&\ + \frac{g_W}{2 c_W} X_{tq}\ \bar{t}\gamma_{\mu}(x^{L}P_{L}+x^{R}P_{R})q\ Z^{\mu} \bigg]
		+ \text{h.c.} \,,
	\end{split}
\end{equation}
%--------------------------------
where $\lambda_{tq}$, $\kappa_{tq}$ and $X_{tq}$ are dimensionless real parameters that denote
the strength of the anomalous FCNC couplings. In the above effective Lagrangian, the chirality parameters
are normalized to $|\lambda^{a}|^2 + |\lambda^{v}|^2$ = $|x^{L}|^2 + |x^{R}|^2$ = $|\kappa^{v}|^2 + |\kappa^{a}|^2$ = 1
and $P_{L,R}$ are the left- and right-handed projection operators, $P_{L,R}=\frac{1}{2} (1 \mp \gamma^5)$.
The anomalous FCNC interactions $tq \gamma$ and $tqZ$ lead to production of a top quark
in association with a light quark in electron-positron collisions.
The Feynman diagram for this process is shown in Figure~\ref{feynman-diagram} including the subsequent leptonic decay of the W boson in the top quark decay.
In Table~\ref{cross-sections1}, the cross sections of $e^{-} + e^{+} \rightarrow t\bar{u} + t\bar{c} + \bar{t}u + \bar{t}c$ including the
branching ratio of the top quark decays into a $W$ boson and a b-quark,  and $W$ boson decays into
a charged lepton (muon and electron) and a neutrino are presented. The cross sections are
shown at three different center-of-mass energies of 240, 350 and 500 GeV.
It should be pointed out that the cross sections due to photon and $Z$ boson
exchange are different and depends on the type of FCNC coupling.
The contribution of photon and $Z$ boson exchange with the $\sigma^{\mu \nu}$ coupling
increases with the energy of the center-of-mass. This is because of the
presence of an additional momentum factor $q^{\nu}$ in the effective Lagrangian.

\begin{figure}[tbh]
	\centering
	\includegraphics[width=0.60\textwidth]{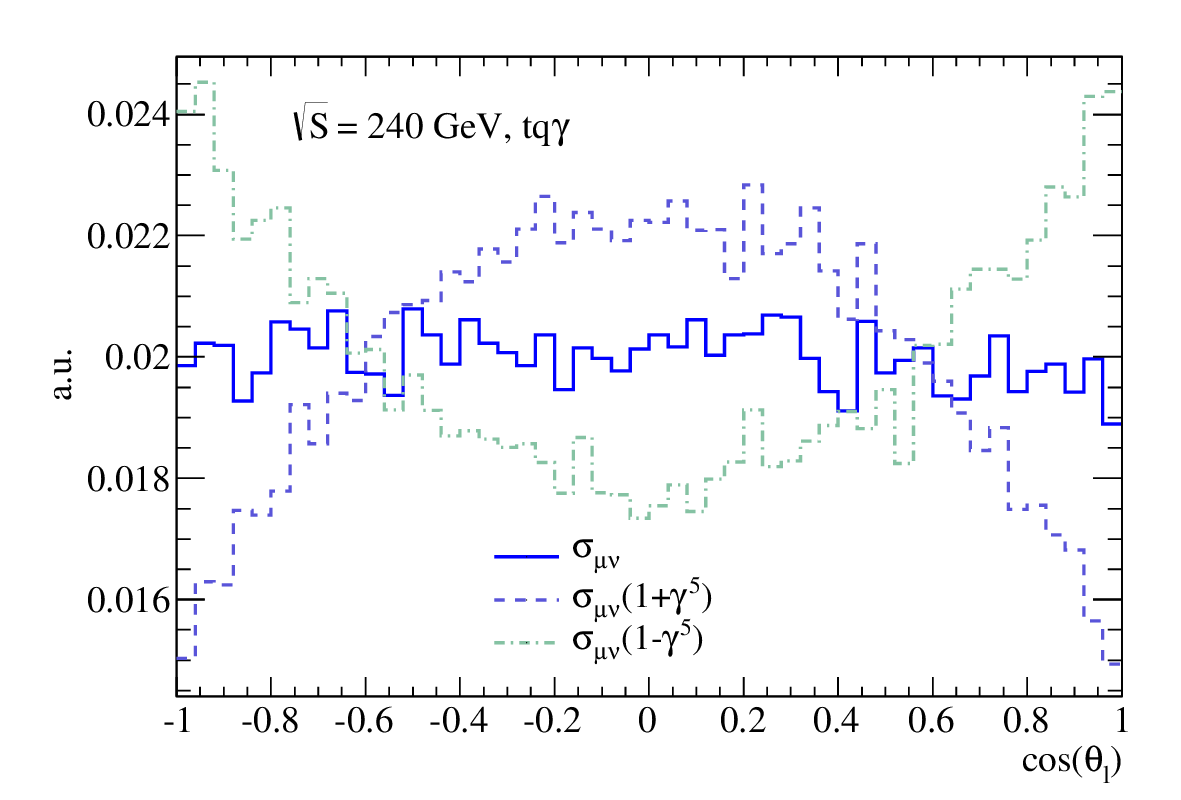}
	\caption{The distribution of the cosine of the angle between the outgoing charged lepton with the $z-$axis for $tq \gamma$ with different chirality
		assumptions at the center-of-mass energy of 240 GeV.}
	\label{cos}
\end{figure}

According to the three independent terms of the Lagrangian, there are three separate ways to produce single top quark
plus a light quark. In this analysis, all three terms of the Lagrangian are investigated independently with the
following sets of the chirality parameters: $\lambda^{v}=1, \lambda^{a}=0$ for $tq \gamma$, for vector like coupling
of $tqZ$: $x^{L}=x^{R}$ while for tensor FCNC coupling of $tqZ$: $\kappa^{v}=1, \kappa^{a}=0$.
In case of observing an excess indicating FCNC signal, the angular distribution of the outgoing particles
can be used to determine the chirality of the FCNC couplings. In Figure~\ref{cos}, the distributions
of the cosine of the angle between the outgoing charged lepton with respect to the $z-$axis (beam axis)
are depicted for the $tq \gamma$ signal scenario with three independent types of couplings: $(\lambda^{v}=1, \lambda^{a}=0)$,
$(\lambda^{v}=1, \lambda^{a}=1)$ and $(\lambda^{v}=1, \lambda^{a}=-1)$ at $\sqrt{s} = 240$ GeV. As it can be seen, for the type
of coupling with no $\gamma^5$ the angular distribution is quite flat while for the type
of coupling with projection operator $1 \pm \gamma^5$, the distribution has a behavior like a parabola with opposite shapes depending on the sign of $\gamma^5$.

%======================
%======================

%======================
%======================

%======================
\begin{table}[tbh]
	\begin{center}
		\begin{tabular}{c|c|c|c}
			$\sqrt {s} $                 & 240 GeV        & 350 GeV         & 500 GeV  \\ \hline \hline
			FCNC coupling               & $\sigma$(fb)     &    $\sigma$(fb)      &    $\sigma$(fb)                 \\  \hline \hline 
			$tq\gamma $                & $2154(\lambda_{tq})^{2}$    & $3832(\lambda_{tq})^{2}$      &    $4302(\lambda_{tq})^{2}$           \\
			$tqZ$    $(\sigma_{\mu\nu})$ & $1434(\kappa_{tq})^{2}$    & $2160(\kappa_{tq})^{2}$      &     $2282(\kappa_{tq})^{2}$         \\
			$tqZ$    $(\gamma_{\mu})$  &$916(X_{tq})^{2}$        & $786(X_{tq})^{2}$           &      $464(X_{tq})^{2}$                    \\  \hline  \hline
		\end{tabular}
	\end{center}
	\caption{ Cross-sections (in fb) of $\sigma(e^{-}+e^{+}\rightarrow t\bar{u}+t\bar{c}+\bar{t}u+\bar{t}c)
		\times Br(t\rightarrow Wb \rightarrow l\nu b)$
		with $\ell = e, \mu$ for three signal scenarios, $tq\gamma $, $tqZ$ (vector-tensor) before applying any cut.  }
	\label{cross-sections1}
\end{table}
%======================

%
%%%%%%%%%%%%%%%%%%%%%%%%%%%%%%%%%%%%%%%%%%   Analysis strategy   %%%%%%%%%%%%%%%%%%%%%%%%%%%%%%%%%%%%%%%%%%%%%%%%%
%
\section{Analysis strategy}\label{Monte-Carlo}

As we have mentioned before, this study is dedicated to probe the $tq \gamma$ and $tqZ$ FCNC couplings via single top quark production at FCC-ee.
The results will be presented at different center-of-mass energies of the colliding electron-positron.
In this section, the details of the event generation and Monte Carlo simulation for signal and backgrounds, event selection, and multivariate
analysis to separate signal process from SM background processes will be presented.

%
%%%%%%%%%%%%%%%%%%%%%%%%%%%%%%%%%%%%%%%%%%   Event generation and simulation    %%%%%%%%%%%%%%%%%%%%%%%%%%%%%%%%%%%%%%%%%%%%%%%%%
%
\subsection{Event generation and simulation}

The signal process is defined as $e^- e^+ \rightarrow Z/\gamma \rightarrow t \bar{q} ~ (\bar{t} q)$, where $q$ is an up or a charm quark.
The top quark decays through SM, $t \rightarrow W^+ b \rightarrow \ell^+ \nu_\ell b$ and $\bar{t} \rightarrow W^- \bar{b} \rightarrow \ell^- \overline\nu_\ell \bar{b}$.
Therefore, the final state consists of a charged lepton, missing energy, a $b$-jet and a light jet.

In order to simulate and generate the signal events, the effective Lagrangian describing the FCNC couplings is implemented
with the \feynrules \ package~\cite{Alloul:2013bka,Christensen:2010wz,Fuks:2012im,Duhr:2011se,Christensen:2009jx}, then the model
has been imported to a UFO module~\cite{Degrande:2011ua} and inserted in \madgraph \ 5 package~\cite{Alwall:2011uj,Alwall:2014hca}.

Based on the expected signature of the signal process, the main background contribution is originating from $W^\pm jj$ production
when the $W$ boson decays leptonically, i.e. $ e^+ e^- \rightarrow W^\pm jj \rightarrow  \ell^+ \nu_\ell j j ( \ell^- \overline\nu_\ell j j )$. 
Other backgrounds to the signal include the $t \bar{t}$ events in semi-leptonic channel and $Z \ell^\pm \ell^\pm$ (with hadronic decay of the $Z$ boson).
All of these backgrounds are generated at leading order with \madgraph \ 5. 
The cross sections of the background processes at the center-of-mass energies of $\sqrt{s}$ = 240, 350 and 500 GeV are
presented in the first row of Tables \ref{Cuts-240}, \ref{Cuts-350}, \ref{Cuts-500}.

We employ \pythia~8.1 package~\cite{Skands:2014pea,Kong:2012vg,Guillaud:2000xga,Sjostrand:2006za}
for parton showering, hadronization and decay of unstable particles.
To reconstruct jets the {\sc FastJet} package~\cite{Cacciari:2011ma,Cacciari:2006sm,Cacciari:2005hq} with an anti-$k_{t}$ algorithm~\cite{Catani:1993hr,Ellis:1993tq} with a cone size of
$ 0.5$ is used. Then the {\sc delphes} \ 3 package~\cite{deFavereau:2013fsa,Mertens:2015kba} is employed to model the detector performance.  
We present the results with 70\% for the efficiency of $b$-tagging, a mistagging rate of 10\% for charm-quark jets and 1\% for other light-flavor jets.
It will be shown that the $b$-tagging requirement plays an important role to reject the background contributions, in particular, 
$W^\pm jj$ and $Z \ell^\pm \ell^\pm$.

The jet energies are smeared in {\sc delphes} similar to an ILD-like detector~\cite{Baer:2013cma,Baer:2013vqa,Barklow:2015tja}
%--------------------------------
\begin{eqnarray}\label{Gaussian-smearing-jets}
	\frac{\Delta  E_{j}}{E_{j}}  =  \frac{30 \%}{\sqrt{E_{j} \ ({\rm GeV})}}   \,,
\end{eqnarray}
%--------------------------------
The detector performance modeling of leptons (electrons and muons) is taken similar to a CMS-like detector which has been described 
in Ref.~\cite{Ball:2007zza}.

\begin{figure*}[tbh]
	\begin{center}
		\includegraphics[width=0.70\textwidth]{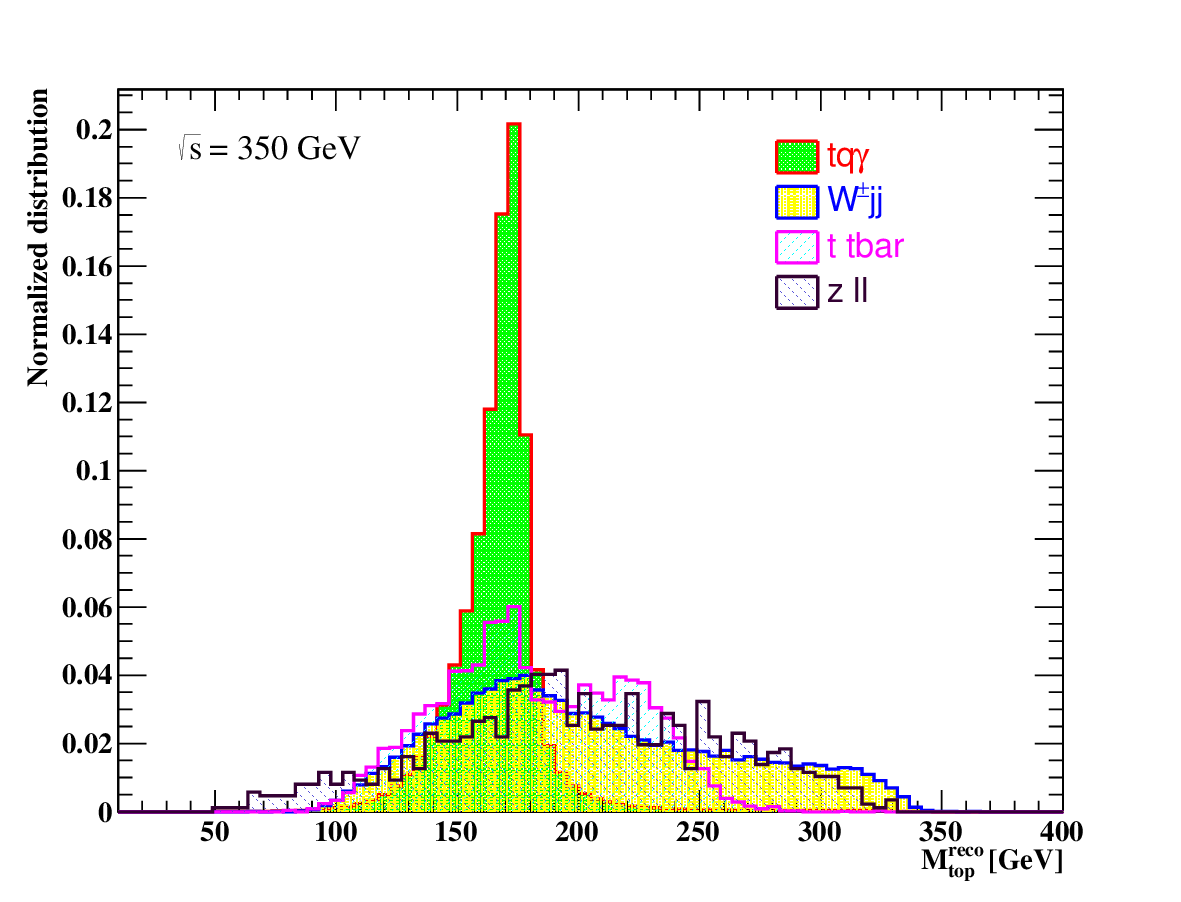}
		\caption{The normalized reconstructed top quark mass distributions for signal ($tq \gamma$) and the corresponding $W^\pm jj$, $t \bar{t}$ and  $Z \ell^\pm \ell^\pm$ SM background processes at $\sqrt{s} = 350$ GeV. The signal has been shown with $\lambda_{tq}=0.1$.}
		\label{reconstructed-top-quark}
	\end{center}
\end{figure*}

Events are preselected by requiring only one charged lepton (electron or muon) 
with $p_T^{\ell} \geq$ 10~${\rm GeV}$ and $|\eta^{\ell}|<$ 2.5.  
No specific requirement is applied as for trigger condition however the presence of an energetic charged lepton is assumed to be enough.
We require to have at least two jets in each event with $p_T^{jet} \geq 10~{\rm GeV}$ and $|\eta^{jet}|<2.5$, from which one is required to be originating from the 
hadronization of a $b$-quark. The events are rejected in which the charged leptons are not isolated. 
In this analysis, the events with at least one $b$-tagged jet are kept and the total number of jets is required to be greater than two.
These requirements with the preliminary requirement of the presence of only one isolated charged lepton
in the final state help suppress the contribution of background events from the top quark pair production.
Among the non-b-tagged jets, the one with highest transverse momentum is chosen to be originating 
from the light quark in the final state.
The four-momentum of the neutrino is determined without any ambiguity from the missing momentum of the event. The missing momentum is 
required be greater than 10 GeV.
To reconstruct the signal topology, first the $W$ boson momentum is reconstructed from the
momenta of the charged lepton and the neutrino as $p_{W} = p_{l}+p_{\nu}$.
The top quark four-momentum is obtained by combining the reconstructed $W$ boson with the $b$-tagged jet. 
The mass distribution of the reconstructed top quark is illustrated in Figure~\ref{reconstructed-top-quark} for
$tq \gamma$ signal and for background processes $W^{\pm}jj$, $t \bar{t}$ and $Z \ell^\pm \ell^\pm$.

%======================
%======================

The distribution is at the center-of-mass energy of 350 GeV. As expected the reconstructed top quark mass distribution for signal 
has a peak around the top quark mass while the background processes have an almost flat distribution with no sharp peak.
The top quark pair background process also has an almost sharp peak on the top quark mass due to the fact that the charged lepton, neutrino, and $b$-jet are coming from one of the top quarks.  The $W^{\pm}jj$ background has a broad invariant mass distribution because the $b$-jet candidate is not originating from the decay of a top quark.

%
%%%%%%%%%%%%%%%%%%%%%%%%%%%%%%%%%%%%%%%%%%    Separation of signal from background    %%%%%%%%%%%%%%%%%%%%%%%%%%%%%%%%%%%%%%%%%%%%%%%%%
%
\subsection{Separation of signal from background}\label{Cut-based-TMA}

In order to reduce the SM background processes which have different topologies from the signal
events, a multivariate technique~\cite{Hocker:2007ht,Stelzer:2008zz,Therhaag:2009dp,Speckmayer:2010zz,Therhaag:2010zz} is used.
After the preselection cuts described in the previous section which consists of the detector acceptance cuts,
and including the effects of $b$-tagging and mistagging, around 40--45\% of the signal
events and 1--4\% of background events are survived. The cross sections of
signal in all scenarios and the corresponding SM backgrounds at three center-of-mass energies after the preselection
cuts are presented in Table~\ref{Cuts-240},~\ref{Cuts-350} and ~\ref{Cuts-500} for $\sqrt {s} $ = 240, 350 and 500 GeV, respectively. 

%======================
\begin{table*}[htbp]
	\begin{center}
		\begin{tabular}{c|ccc|cc}
			$\sqrt {s} = 240 GeV $                                     &  &  \multicolumn{2}{c|}{ Signal }                  & \multicolumn{2}{c}{ Background }  \\ \hline
			Cuts           & $tq\gamma $   &   $tqZ$ $(\sigma_{\mu\nu})$   &    $tqZ$ $(\gamma_{\mu})$  &  $W^\pm jj$ &  $Z \ell^\pm \ell^\pm$  \\
			\hline  \hline
			Cross-sections (in fb)   &  $2154.0(\lambda_{tq})^{2}$  & $1434.0(\kappa_{tq})^{2}$ &  $916.0(X_{tq})^{2}$   &   $4881.2$    &   $3588.4$    \\
			1$\ell$+$|\eta^{\ell}|<2.5$+$P_T^{\ell}>10$+$|\vec{p}_{\rm miss}|>10$  &  $1679.8(\lambda_{tq})^{2}$  & $1117.8(\kappa_{tq})^{2}$ &  $715.6(X_{tq})^{2}$   &   $3886.3$    &    $100.1$    \\
			$\geq2jets$+$|\eta^{jets}|<2.5$+$P_T^{jets}>10$             &  $1393.3(\lambda_{tq})^{2}$  & $927.3(\kappa_{tq})^{2}$  &  $590.9(X_{tq})^{2}$   &   $3459.1$  &   $59.7$    \\
			$n_{b-jet}\geq1$                                            &  $900.5(\lambda_{tq})^{2}$   & $598.7(\kappa_{tq})^{2}$  &  $381.8(X_{tq})^{2}$   &   $185.3$  &   $15.3$    \\ \hline  \hline
		\end{tabular}
	\end{center}
	\caption{Cross-sections (in fb) for the three signal scenarios, $tq \gamma$, $tqZ$ (vector and tensor) and the corresponding $W^\pm jj$ and  $Z \ell^\pm \ell^\pm$ SM backgrounds passing sequential cuts at  $\sqrt {s} $ = 240 GeV.}
	\label{Cuts-240}
	%\end{table*}
    %======================
	%\begin{table}[htbp]
	\begin{center}
		\begin{tabular}{c|ccc|ccc}
			$\sqrt {s} = 350 GeV $                                     &  &  \multicolumn{2}{c|}{ Signal }                  & \multicolumn{2}{c}{ Background }  \\ \hline
			Cuts           & $tq\gamma $   &   $tqZ$ $(\sigma_{\mu\nu})$   &    $tqZ$ $(\gamma_{\mu})$  &  $W^\pm jj$ & $t \bar{t}$  & $Z \ell^\pm \ell^\pm$  \\
			\hline  \hline
			Cross-sections (in fb)&  $3832.0(\lambda_{tq})^{2}$      & $2160.0(\kappa_{tq})^{2}$ &   $786.0(X_{tq})^{2}$   & $3221.1$  &  $62.53$  &   $4085.0$ \\
			1$\ell$+$|\eta^{\ell}|<2.5$+$P_T^{\ell}>10$+$|\vec{p}_{\rm miss}|>10$      &  $2984.2(\lambda_{tq})^{2}$      & $1680.2(\kappa_{tq})^{2}$ &   $614.6(X_{tq})^{2}$   & $2447.6$  &  $40.5$   &   $129.5$  \\
			$\geq2jets$+$|\eta^{jets}|<2.5$+$P_T^{jets}>10$         &  $2499.1(\lambda_{tq})^{2}$      & $1405.6(\kappa_{tq})^{2}$ &   $507.9(X_{tq})^{2}$   &  $2175.5$ &   $0.65$  &   $77.7$    \\
			$n_{b-jet}\geq1$                 &  $1614.1(\lambda_{tq})^{2}$      & $909.0(\kappa_{tq})^{2}$  &   $328.4(X_{tq})^{2}$   &  $112.4$  &   $0.43$  &   $20.3$     \\ \hline  \hline
		\end{tabular}
	\end{center}
	\caption{Cross-sections (in fb)  for three signal scenarios, $tq \gamma$, $tqZ$ (vector and tensor) and the corresponding $W^\pm jj$, $t \bar{t}$ and  $Z \ell^\pm \ell^\pm$ SM backgrounds passing sequential cuts at  at $\sqrt {s} $ = 350 GeV.}
	\label{Cuts-350}
	%\end{table}
    %======================
	%\begin{table}[htbp]
	\begin{center}
		\begin{tabular}{c|ccc|ccc}
			$\sqrt {s} = 500 GeV $                                     &  &  \multicolumn{2}{c|}{ Signal }                  & \multicolumn{2}{c}{ Background }  \\ \hline
			Cuts           & $tq\gamma $   &   $tqZ$ $(\sigma_{\mu\nu})$   &    $tqZ$ $(\gamma_{\mu})$  &  $W^\pm jj$ & $t \bar{t}$  & $Z \ell^\pm \ell^\pm$  \\
			\hline  \hline
			Cross-sections (in fb)&  $4302.0(\lambda_{tq})^{2}$      & $2282.0(\kappa_{tq})^{2}$ &   $464.0(X_{tq})^{2}$   & $2048.8$  &  $148.7$   &   $4070.0$ \\
			1$\ell$+$|\eta^{\ell}|<2.5$+$P_T^{\ell}>10$+$|\vec{p}_{\rm miss}|>10$      &  $3277.3(\lambda_{tq})^{2}$      & $1736.9(\kappa_{tq})^{2}$ &   $355.1(X_{tq})^{2}$   & $1383.7$  &  $106.9$   &   $144.7$  \\
			$\geq2jets$+$|\eta^{jets}|<2.5$+$P_T^{jets}>10$         &  $2757.8(\lambda_{tq})^{2}$      & $1460.8(\kappa_{tq})^{2}$ &   $292.0(X_{tq})^{2}$   &  $1242.1$ &  $1.38$    &   $89.7$    \\
			$n_{b-jet}\geq1$                 &  $1776.9(\lambda_{tq})^{2}$      & $941.9(\kappa_{tq})^{2}$  &   $188.0(X_{tq})^{2}$   &  $60.9$   &  $0.89$    &   $24.3$    \\ \hline  \hline
		\end{tabular}
	\end{center}
	\caption{Cross-sections (in fb) for the three signal scenarios, $tq \gamma$, $tqZ$ (vector and tensor) and the corresponding $W^\pm jj$, $t \bar{t}$ and  $Z \ell^\pm \ell^\pm$ SM backgrounds passing sequential cuts at  $\sqrt {s}$ = 500 GeV.}
	\label{Cuts-500}
\end{table*}
%======================

\begin{figure*}[htbp]
	\begin{center}
		\includegraphics[width=0.45\textwidth]{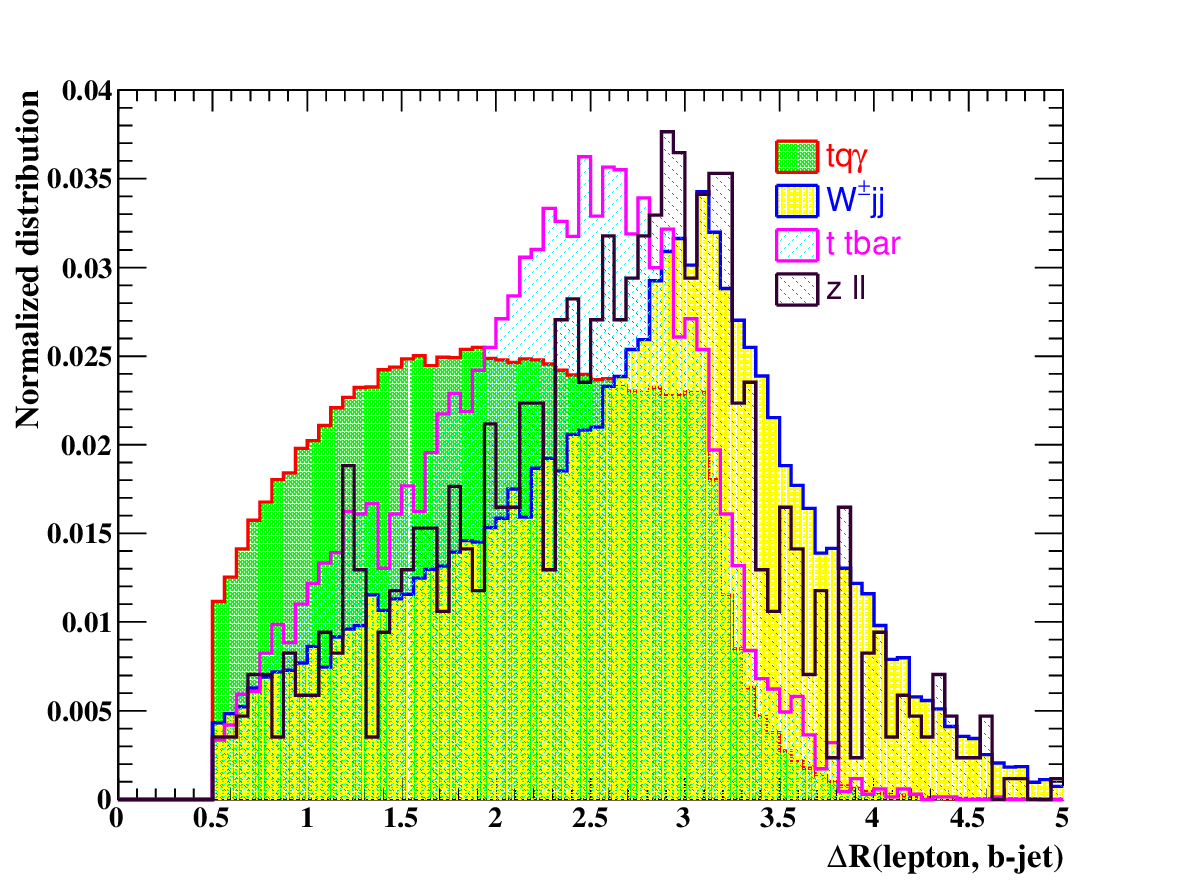}
		\includegraphics[width=0.45\textwidth]{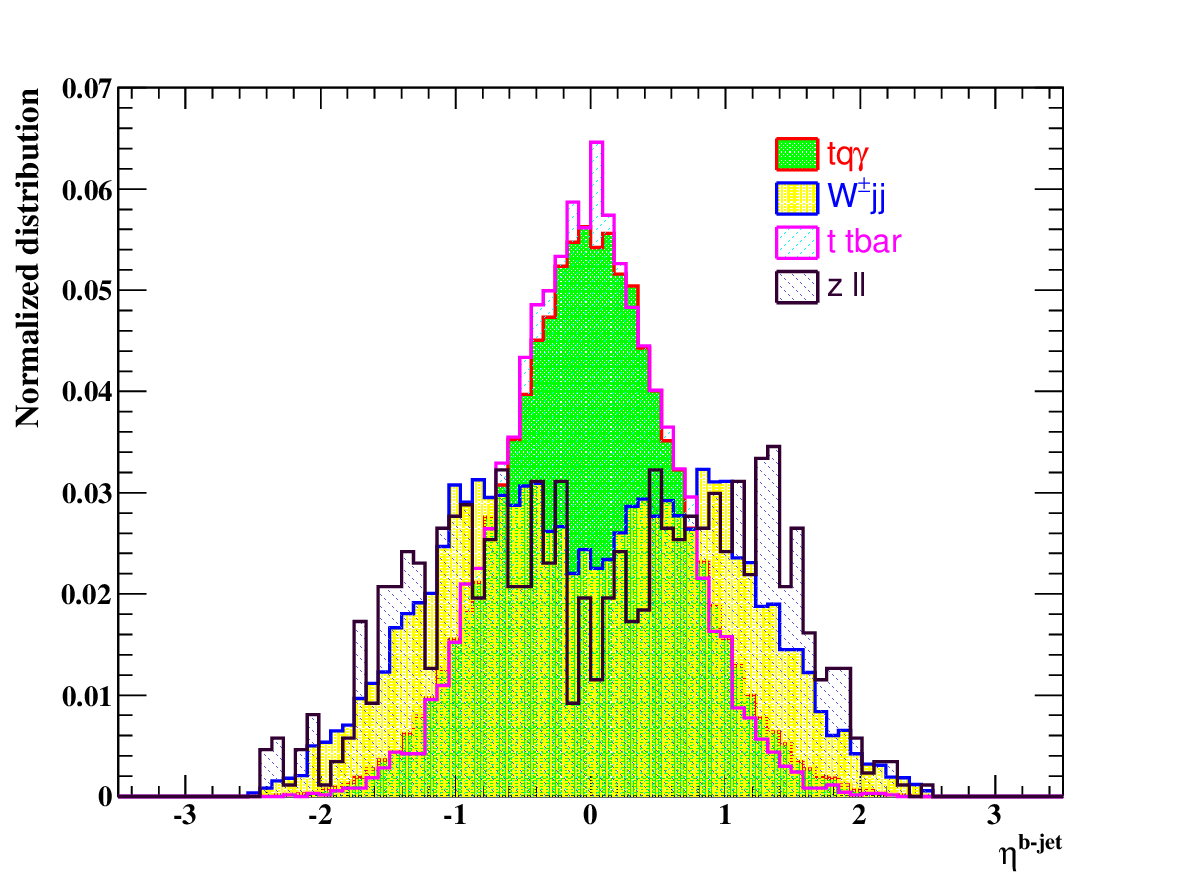}     \\
		\includegraphics[width=0.45\textwidth]{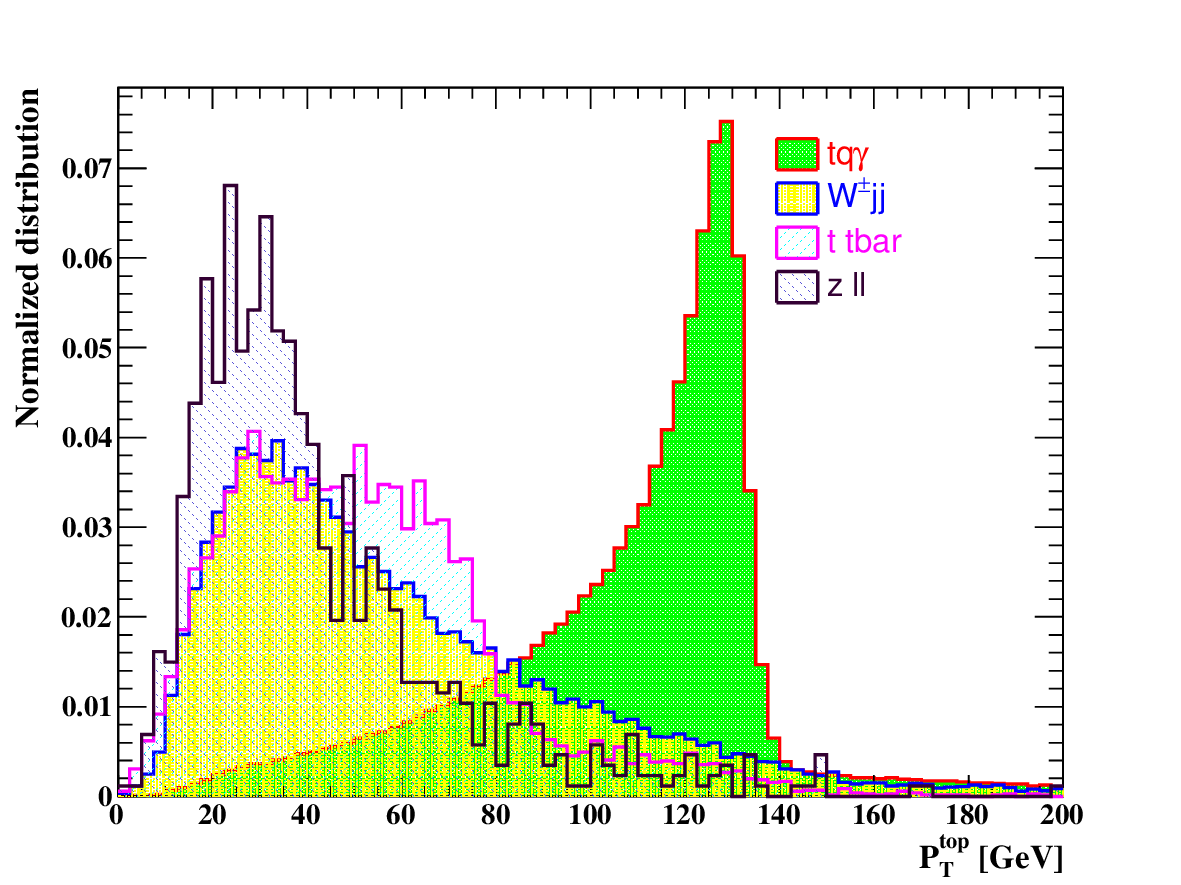}
		\includegraphics[width=0.45\textwidth]{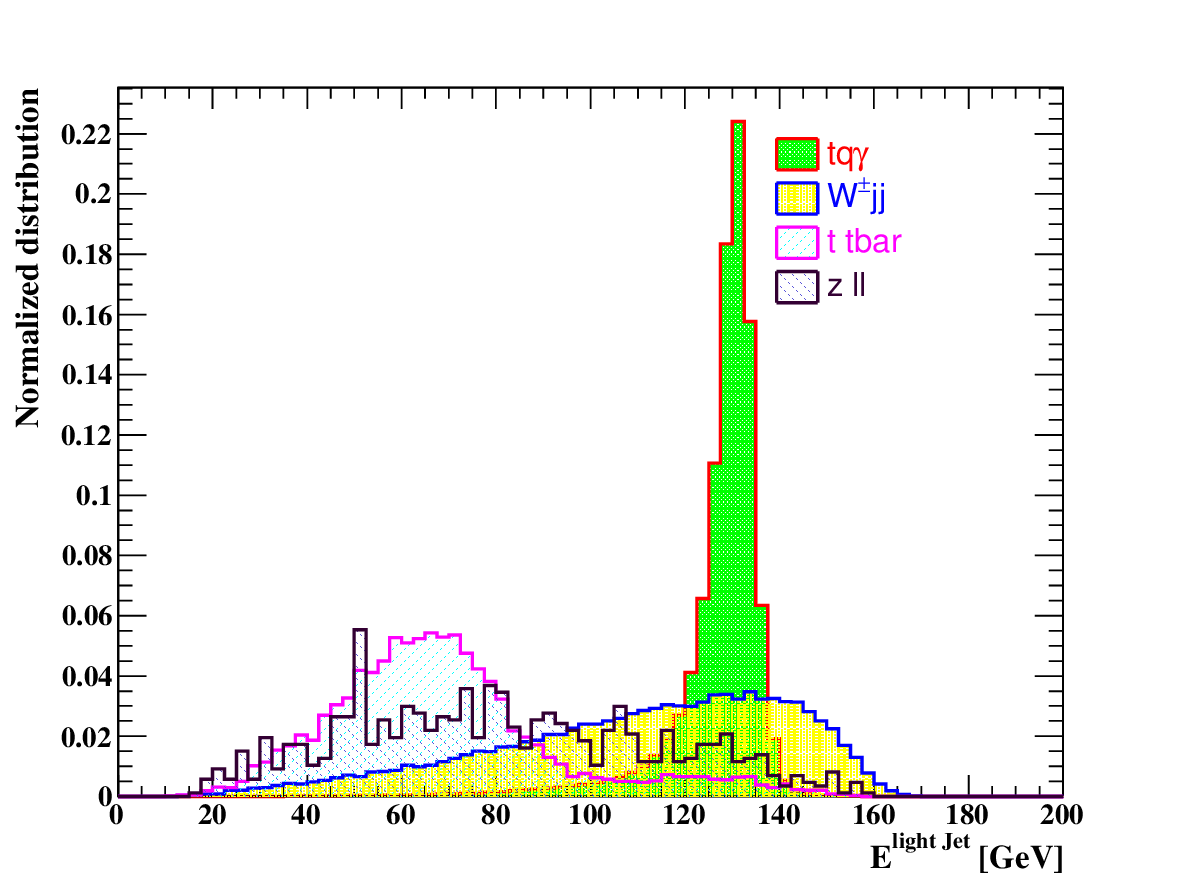}     \\
		\includegraphics[width=0.45\textwidth]{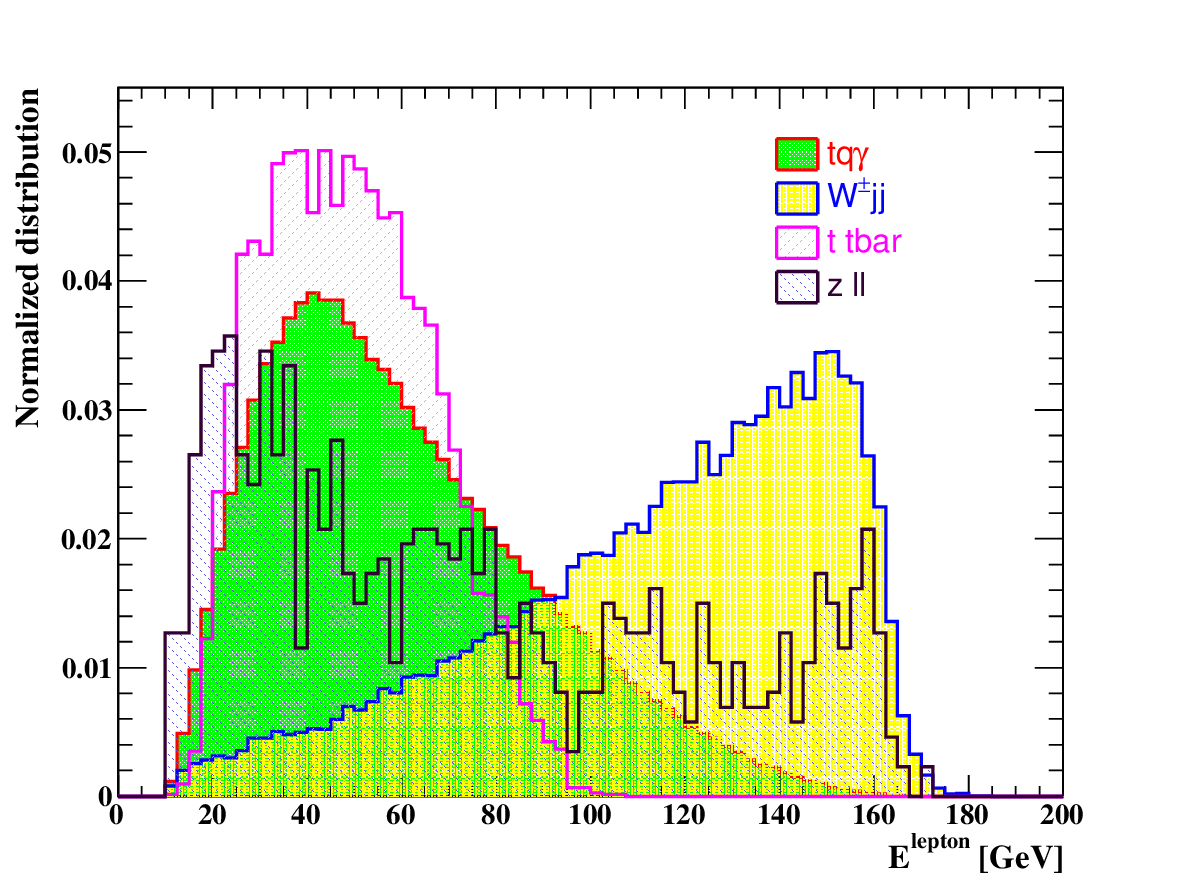}
		\includegraphics[width=0.45\textwidth]{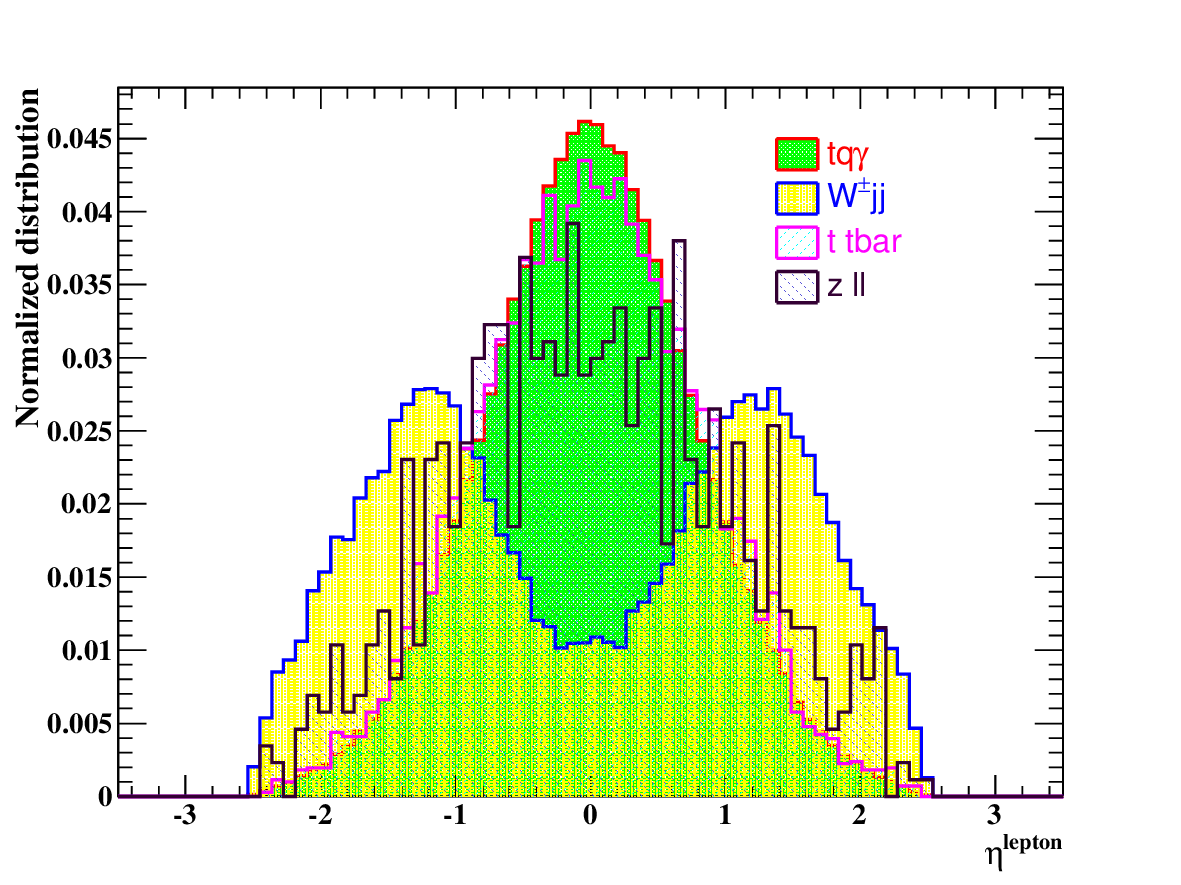}     \\
		\caption{The normalized distributions of some of the input variables to the  multivariate analysis for the
			the center-of-mass energy of 350 GeV.}
		\label{Variables-Distributions}
	\end{center}
\end{figure*}

These preselection cuts are generally loose on a single variable and
remove a large fraction of the background events while barely reducing also the signal
events. In order to obtain a better separation of signal from background events,
a multivariate technique is used. The choice of proper set of variables is important in keeping the signal events, while reducing the large fraction
of SM background events. We select those variables which have
the best possible discrimination power between signal and background processes.
The following variables are used in the analysis:
%-------------------------------------
\begin{itemize}
	\item{$\Delta R(\ell, \rm b-jet)$: the angular separation between the lepton and $b$-jet }
	\item{$p_{T}^{\rm b-jet}, \eta^{\rm b-jet}$: the transverse momentum and pseudorapidity of the $b$-jet}
	\item{$M^{\rm rec}_{\rm top}$: the reconstructed top quark mass}
	\item{$E^\ell,\eta^{\ell}$: the energy and pseudorapidity of the charged lepton}
	\item{$P_T^{\rm top}$: the transverse momentum of reconstructed top-quark}
	\item{$E^{\rm light-jet}$: the energy of the light jet}
\end{itemize}
%-------------------------------------
The distributions of some of these variables are shown
in Figure~\ref{Variables-Distributions}. These distributions are corresponding to
the signal scenario with anomalous $tq \gamma$ coupling at the center-of-mass energy of 350 GeV.
For all signal scenarios $tq \gamma$, $tqZ(\gamma^{\mu})$ and $tqZ(\sigma_{\mu\nu})$
the same variables are used as the inputs of the multivariate analyses. The analyses are performed separately at the
center-of-mass energies of 240, 350 and 500 GeV. Going to higher center-of-mass energies leads to reduce the overlapping between the signal and background distributions in the MVA output.
 
%======================
%======================

The cross sections of the signal and the $W^\pm jj$, $t \bar{t}$ and $Z \ell^\pm \ell^\pm$ background processes 
after performing the multivariate analysis are presented in
Table~\ref{cross-section3}. As can be seen from this table, the background rejection
rate varies at different center-of-mass energies. For all signal scenarios, the background
rejection rates after the multivariate analysis technique are $\sim 10^{-1}$, $\sim 10^{-2}$ and $\sim 10^{-3}$ at the center-of-mass
energies of 240, 350, 500 GeV, respectively. The discriminating power of the input
variables are increasing with the center-of-mass energies of the collision.
Going to higher energies the overlapping between the signal and background
distributions is reduced. In particular, this happens for the top mass, lepton energy
and the top quark transverse momentum distributions. Larger background suppression
is achieved for the $tqZ$ signal with $\sigma_{\mu\nu}$ coupling with respect to the $\gamma_{\mu}$ coupling.
Since the signal-to-background ratio for all signal scenarios increases
with the increment of the center-of-mass energy, more sensitivity is expected at larger energies.

%======================
\begin{table*}[htbp]
	\begin{center}
		\begin{tabular}{l|c|c c c c c }  \hline \hline
			$\sqrt {s}$=240 GeV                      & Couplings & Signal &  ~  $W^\pm jj$ (fb)                    ~  &              &  ~  $Z \ell^\pm \ell^\pm$ (fb)               \\  \hline
			& $t q \gamma$                       & $826.32(\lambda_{tq})^{2}$ &  ~  $26.59$  ~  & -  &  ~  $5.27$      \\
			TMVA & $t q Z$ $(\sigma_{\mu \nu})$  & $547.90(\kappa_{tq})^{2}$ &  ~  $25.65$  ~  & -  & ~   $2.15$      \\
			& $t q Z$ $(\gamma_{\mu})$           & $354.13(X_{tq})^{2}$  &  ~  $30.56$  ~  & -  &  ~  $2.57$      \\   \hline  \hline
			$\sqrt {s}$=350 GeV                      &  &  Signal  &  ~  $W^\pm jj$ (fb)                  ~  &   $t \bar{t}$ (fb)       &  ~  $Z \ell^\pm \ell^\pm$ (fb)           \\  \hline
			& $t q \gamma$                       & $1521.31(\lambda_{tq})^{2}$ &  ~  $7.59$  ~  &  $0.034$ &  ~  $1.45$      \\
			TMVA & $t q Z$ $(\sigma_{\mu \nu})$  & $856.72(\kappa_{tq})^{2}$ &  ~  $7.61$  ~  &  $0.031$ &  ~  $1.45$      \\
			& $t q Z$ $(\gamma_{\mu})$           & $306.48(X_{tq})^{2}$ &  ~  $8.49$  ~  &  $0.37$  &  ~  $1.74$      \\   \hline  \hline
			$\sqrt {s}$=500 GeV                    &    & Signal  &  ~  $W^\pm jj$ (fb)                   ~  &   $t \bar{t}$  (fb)         &  ~  $Z \ell^\pm \ell^\pm$ (fb)         \\  \hline
			& $t q \gamma$                      & $1677.29(\lambda_{tq})^{2}$  &  ~  $2.11$  ~  &  $0.11$ &  ~  $0.64$      \\
			TMVA & $t q Z$ $(\sigma_{\mu \nu})$ & $895.71(\kappa_{tq})^{2}$  &  ~  $2.43$  ~  &  $0.14$ &  ~  $0.64$      \\
			& $t q Z$ $(\gamma_{\mu})$          & $176.68(X_{tq})^{2}$  &  ~  $3.07$  ~  &  $0.13$ &  ~  $1.21$      \\   \hline  \hline
		\end{tabular}
	\end{center}
	\caption{ Cross-sections (in fb) of signal and $W^\pm jj$, $t \bar{t}$ and $Z \ell^\pm \ell^\pm$ background processes after performing the multivariate analysis for three signal scenarios, $tq \gamma $, $tqZ$ (vector and tensor) at $\sqrt {s} = 240$, $350$ and $500$ GeV .}
	\label{cross-section3}
\end{table*}
%======================

%
%%%%%%%%%%%%%%%%%%%%%%%%%%%%%%%%%%%%%%%%%%    Sensitivity estimation    %%%%%%%%%%%%%%%%%%%%%%%%%%%%%%%%%%%%%%%%%%%%%%%%%
%
\section{Sensitivity estimation} \label{results}

\begin{figure*}[htbp]
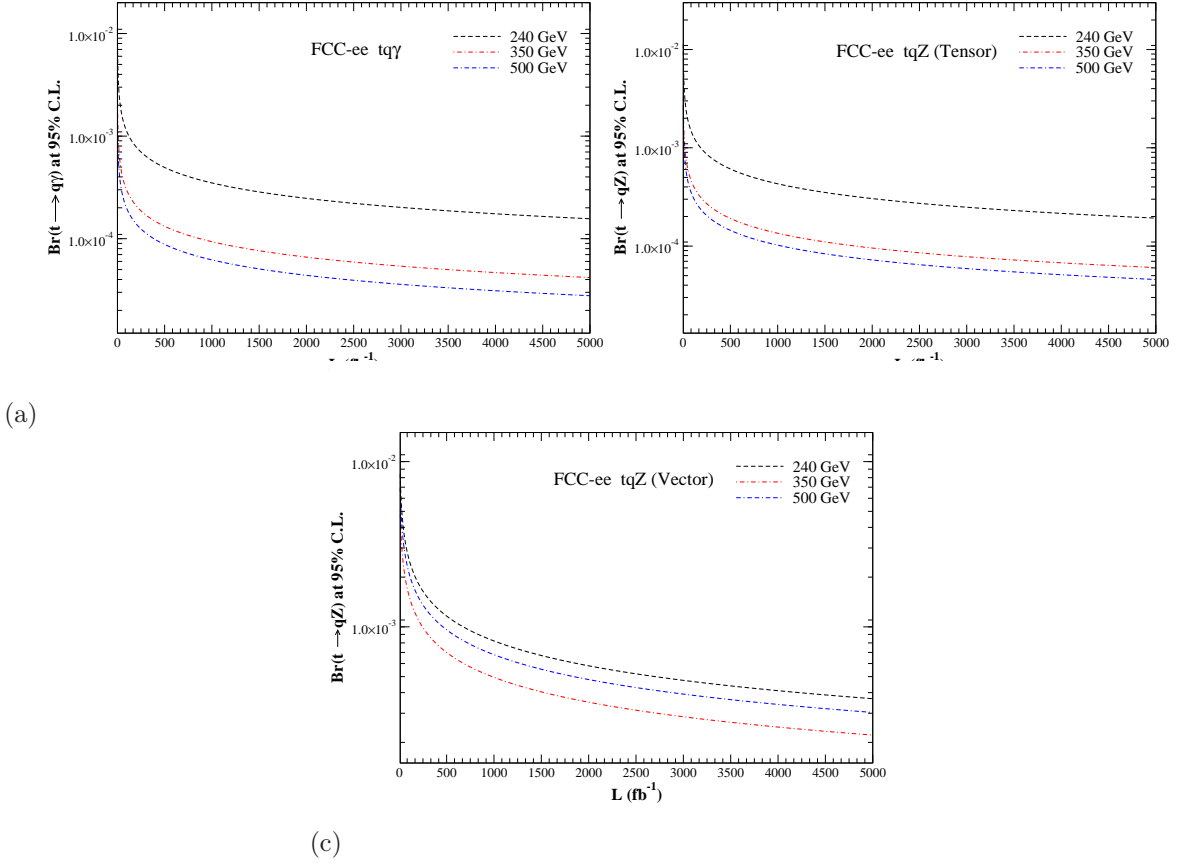

	\begin{center}
		\includegraphics[width=0.45\textwidth]{BrtqALuminosity.eps}
		\includegraphics[width=0.45\textwidth]{BrtqZSigmamuLuminosity.eps}     \\
		(a) \hspace*{0.45\textwidth}     (b) \hspace*{0.45\textwidth} \\
		\includegraphics[width=0.45\textwidth]{BrtqZgammamuLuminosity.eps}     \\
		(c) \hspace*{0.45\textwidth} \\
		\caption{Branching ratios of a FCNC signal detectable at the $3 \sigma$ level as a function of
			integrated luminosity at $\sqrt{s}$ = 240, 350 and 500 GeV of FCC-ee energies.
			(a) for $ Br (t \to q \gamma )$,
			(b) for $ Br (t \to q Z)$ $(\sigma_{\mu\nu})$,
			and (c) for $Br (t \to q Z)$  $(\gamma_{\mu})$.}
		\label{Figure:BranchingRatio}
	\end{center}
\end{figure*}

To estimate the sensitivities, the expected $3 \sigma$ significance and the
upper limits on the  branching ratios at 95\% C.L are presented.
The $3 \sigma$ discovery ranges are obtained using the significance $S/\sqrt{B}$, where
$S$ and $B$ are the number of signal and background events after all selections, respectively.
Without including any systematic effects, the $3 \sigma$ discovery regions of the branching ratios are
presented in Table~\ref{3sigma-500} for three signal scenarios at three center-of-mass energies. 
The $3 \sigma$ discovery regions in terms of the integrated luminosity are
also depicted in Figure~\ref{Figure:BranchingRatio}. We observe that at $3 \sigma$ significance level
branching ratios at the order of $10^{-3}-10^{-4}$ is achievable at the center-of-mass energy of 240 GeV
while going to larger energies of 350 and 500 GeV can lead to an improvement of
one order of magnitude for $tq \gamma$ and $tqZ(\sigma_{\mu\nu})$ with an integrated luminosity of 300 fb$^{-1}$.
The FCNC transition of $t \rightarrow qZ$ with $\gamma_{\mu}$-type couplings would not
be measured better than $10^{-4}$. 
According to Figure~\ref{Figure:BranchingRatio}, going to high luminosity regime
at the center-of-mass energies of 240 and 350 leads to a reach sensitivity at the
order of $10^{-5}$. 

%======================
\begin{table*}[htbp]
	\begin{center}
		\begin{tabular}{c| c c c c }  \hline
			Integrated luminosity & Branching ratio     &  ~  240 GeV          ~  &   350 GeV        &  ~  500 GeV           \\  \hline  \hline
			& $Br (t \to q \gamma )$                    &  ~  $6.38 \times 10^{-4}$  ~  &  $1.70 \times 10^{-4}$ &  ~  $1.13 \times 10^{-4}$      \\
			300 fb$^{-1}$& $Br (t \to q Z)$  $(\sigma_{\mu \nu})$    &  ~  $7.85 \times 10^{-4}$  ~  &  $2.46 \times 10^{-4}$ &  ~  $1.86 \times 10^{-4}$      \\
			& $Br (t \to q Z)$  $(\gamma_{\mu})$        &  ~  $1.50 \times 10^{-3}$  ~  &  $9.03 \times 10^{-4}$ &  ~  $1.23 \times 10^{-3}$      \\   \hline  \hline
			&$Br (t \to q \gamma )$                    &  ~  $2.01 \times 10^{-4}$  ~  &  $5.39 \times 10^{-5}$ &  ~  $3.58 \times 10^{-5}$      \\
			3 ab$^{-1}$& $Br (t \to q Z)$  $(\sigma_{\mu \nu})$    &  ~  $2.48 \times 10^{-4}$  ~  &  $7.79 \times 10^{-5}$ &  ~  $5.90 \times 10^{-5}$      \\
			& $Br (t \to q Z)$  $(\gamma_{\mu})$        &  ~  $4.73 \times 10^{-4}$  ~  &  $2.85 \times 10^{-4}$ &  ~  $3.91 \times 10^{-4}$      \\   \hline  \hline
			& $Br (t \to q \gamma )$                    &  ~  $2.01 \times 10^{-5}$  ~  &  $2.95 \times 10^{-5}$ &  ~  $1.96 \times 10^{-5}$      \\
			10 ab$^{-1}$& $Br (t \to q Z)$  $(\sigma_{\mu \nu})$    &  ~  $2.44 \times 10^{-5}$  ~  &  $4.27 \times 10^{-5}$ &  ~  $3.23 \times 10^{-5}$      \\
			& $Br (t \to q Z)$  $(\gamma_{\mu})$        &  ~  $2.59 \times 10^{-4}$  ~  &  $1.56 \times 10^{-4}$ &  ~  $2.14 \times 10^{-4}$      \\   \hline  \hline
		\end{tabular}
	\end{center}
	\caption{The sensitivity for a significance level of $3 \sigma$  at the center-of-mass energies of 240, 350, 500 GeV for the integrated luminosities of 300 fb$^{-1}$, 3 ab$^{-1}$, 10 ab$^{-1}$.}\label{3sigma-500}
\end{table*}
%======================

%======================
%======================

In order to set 95\% CL upper limits on the anomalous FCNC couplings and consequently on the branching ratios, the CL$s$ technique is used~\cite{Read:2002hq}.
For the limits calculations the RooStats~\cite{Moneta:2010pm} package is used.
The 95\% C.L upper limits on the branching ratios of $t \rightarrow q\gamma$ and $t \rightarrow qZ$ at the center-of-mass energies of 240, 350 and 500 GeV
are shown in Table~\ref{cls-500} based on an integrated luminosity of 300 fb$^{-1}$, 3 ab$^{-1}$ and 10 ab$^{-1}$. 
As we expected, at each center-of-mass energy, the loosest limits belong to the FCNC transition of $t \rightarrow qZ$ with $\gamma_{\mu}-$type
coupling ($10^{-4}$). We note that the larger center-of-mass energy leads to  even  one order of magnitude tighter bounds. 

%======================
\begin{table*}[htbp]
	\begin{center}
		\begin{tabular}{c| c c c c }  \hline
			Integrated luminosity & Branching ratio     &  ~  240 GeV          ~  &   350 GeV        &  ~  500 GeV           \\  \hline  \hline
			&$Br (t \to q \gamma )$                    &  ~  $1.23 \times 10^{-4}$  ~  &  $3.43 \times 10^{-5}$ &  ~  $2.45 \times 10^{-5}$      \\
			300 fb$^{-1}$&$Br (t \to q Z)$  $(\sigma_{\mu \nu})$    &  ~  $1.50 \times 10^{-4}$  ~  &  $4.97 \times 10^{-5}$ &  ~  $3.94 \times 10^{-5}$      \\
			&$Br (t \to q Z)$  $(\gamma_{\mu})$        &  ~  $3.06 \times 10^{-4}$  ~  &  $1.83 \times 10^{-4}$ &  ~  $2.67 \times 10^{-4}$      \\   \hline  \hline
			&$Br (t \to q \gamma )$                    &  ~  $3.70 \times 10^{-5}$  ~  &  $9.86 \times 10^{-6}$ &  ~  $6.76 \times 10^{-6}$      \\
			3 ab$^{-1}$&$Br (t \to q Z)$  $(\sigma_{\mu \nu})$    &  ~  $4.50 \times 10^{-5}$  ~  &  $1.41 \times 10^{-5}$ &  ~  $1.09 \times 10^{-5}$      \\
			&$Br (t \to q Z)$  $(\gamma_{\mu})$        &  ~  $9.25 \times 10^{-5}$  ~  &  $5.27 \times 10^{-5}$ &  ~  $7.49 \times 10^{-4}$      \\   \hline  \hline
			&$Br (t \to q \gamma )$                    &  ~  $2.01 \times 10^{-5}$  ~  &  $5.25 \times 10^{-6}$ &  ~  $3.59 \times 10^{-6}$      \\
			10 ab$^{-1}$&$Br (t \to q Z)$  $(\sigma_{\mu \nu})$    &  ~  $2.44 \times 10^{-5}$  ~  &  $7.60 \times 10^{-6}$ &  ~  $5.85 \times 10^{-6}$      \\
			&$Br (t \to q Z)$  $(\gamma_{\mu})$        &  ~  $5.02 \times 10^{-5}$  ~  &  $2.83 \times 10^{-5}$ &  ~  $4.00 \times 10^{-5}$      \\   \hline  \hline
		\end{tabular}
	\end{center}
	\caption{The upper limits on the top FCNC decays at 95\% C.L obtained 
		at the center-of-mass energies of 240, 350 and 500 GeV for the integrated luminosities of 300 fb$^{-1}$, 3 ab$^{-1}$, 10 ab$^{-1}$.}
	\label{cls-500}
\end{table*}
%======================

In order to investigate the sensitivity to $b$-tagging efficiency and miss-tagging rates, we also present the 95\% C.L upper limits on the branching ratios of $t \rightarrow q\gamma$ and $t \rightarrow qZ$  for 85\% of $b$-tagging efficiency and a 5\% mistagging rates. The results correspond to the center-of-mass energy of 350 GeV for the integrated luminosity of 300 fb$^{-1}$. As can be seen from Table~\ref{limits-85}, the higher $b$-tagging efficiency and smaller mistagging rates could improve the  branching ratios by a factor of around 1.6.
Charm-tagging algorithm could leads to distinguish between $tuV$ and $tcV$ FCNC interactions. It is found that a charm tagging algorithm with an efficiency of 30\% provides the possibility to separate $tuV$ and $tcV$ and branching fractions of $t \to c \gamma$ down to $10^{-5}$ with an integrated luminosity of 10 ab$^{-1}$ at the center-of-mass energy of 240 GeV is achievable.

The effect of systematic uncertainties is considered for two assumed values of overall uncertainties: 5\% and 10\%. The change on the branching fraction of $Br(t \to q \gamma)$, $\Delta Br$, are $0.50 \times 10^{-5}$ and  $5.47 \times 10^{-5}$ for the uncertainties of 5\% and 10\%, respectively.

%======================
\begin{table*}[htbp]
	\begin{center}
		\begin{tabular}{l|c c c c }  \hline
			$\sqrt {s}$         &  ~  $Br (t \to q \gamma )$       &  ~ $Br (t \to q Z)$  $(\sigma_{\mu \nu})$  &  ~  $Br (t \to q Z)$  $(\gamma_{\mu})$  \\  \hline
			350 GeV           &  ~  $2.19 \times 10^{-5}$        & ~ $3.12 \times 10^{-5}$                    &  ~  $1.22 \times 10^{-4}$      \\   \hline  \hline
		\end{tabular}
	\end{center}
	\caption{The upper limits on the top FCNC decays at $95\%$ C.L obtained using the CL$s$ method
		at the $\sqrt {s}$ = 350 GeV for $85\%$ of b-tagging efficiency and a $5\%$ mistagging rates based on an integrated luminosity of 300 fb$^{-1}$.}
	\label{limits-85}
\end{table*}
%======================

\begin{figure}[htbp]
	\centering
	\includegraphics[width=0.70\textwidth]{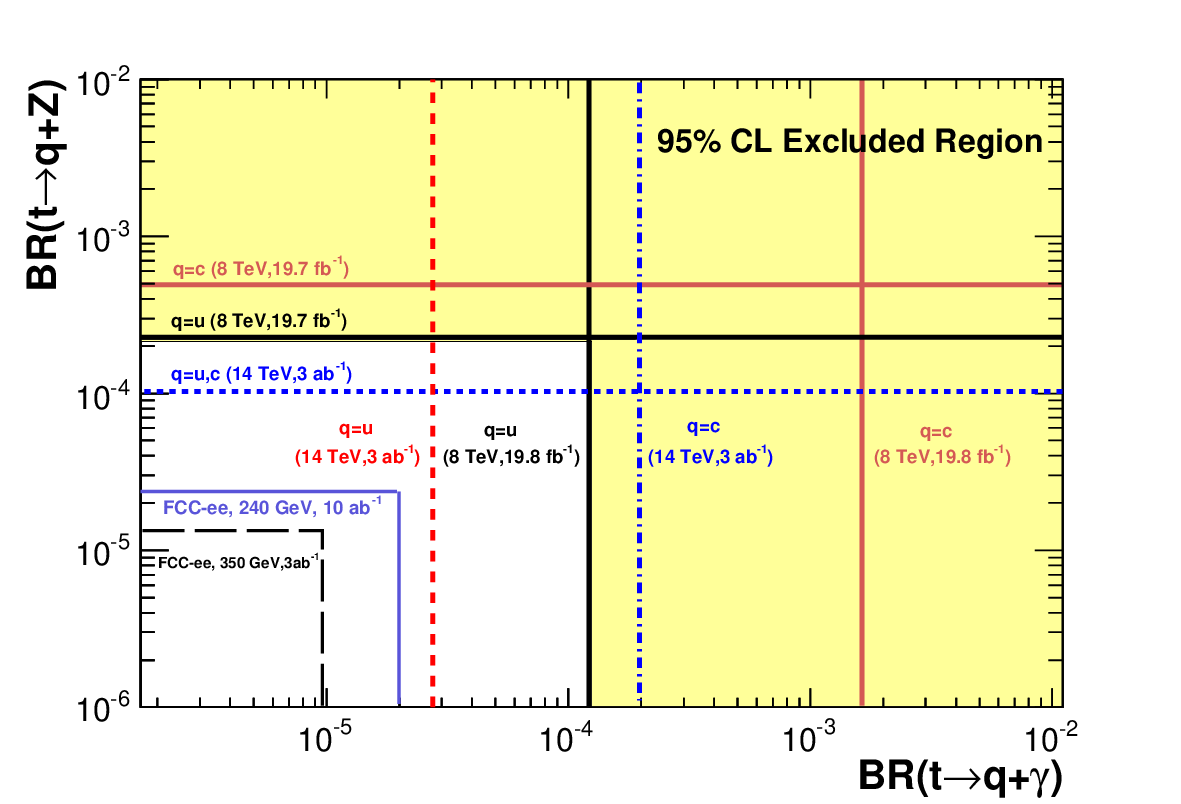}
	\caption{ The current observed upper limits on the $Br(t \rightarrow qZ)$ versus $Br(t \rightarrow q\gamma)$ at 95\% C.L
		from the recent analyses of the CMS experiment~\cite{CMS:2016bss,Khachatryan:2015att}. The expected sensitivity from the CMS experiment with 3000 fb$^{-1}$ is
		also shown~\cite{CMS-HL}. The sensitivity of the FCC-ee with 3 ab$^{-1}$ at the center-of-mass energy of 350 GeV, and with 10 ab$^{-1}$ at the center-of-mass energy of 240 GeV are
		presented as well.}
	\label{brz-bra}
\end{figure}

In Figure~\ref{brz-bra}, we present the current observed upper limits on the $Br(t \rightarrow qZ)$ versus $Br(t \rightarrow q\gamma)$ at 95\% CL
from CMS experiments~\cite{CMS:2016bss,Khachatryan:2015att}. The expected sensitivity from the CMS experiment with 3000 fb$^{-1}$
in proton-proton collisions at the center-of-mass energy of 14 TeV is also shown~\cite{CMS-HL}.
The sensitivity of the FCC-ee with 3 ab$^{-1}$ at the center-of-mass energy of 350 GeV, and with 10 ab$^{-1}$ at the center-of-mass energy of 240 GeV are
compared with the CMS experiment results. With an integrated luminosity of 3000 fb$^{-1}$, CMS is expected to
reach to an upper limit of $2.7 \times 10^{-5}$ on the branching ratio of $t \rightarrow u \gamma$, $\, 2.0 \times 10^{-4}$ on the branching ratio of $t \rightarrow c \gamma$, and $\, 1.0 \times 10^{-4}$ on
the branching ratio of $t \rightarrow qZ$ ($\sigma_{\mu\nu}-$type coupling).
The FCC-ee potential upper limits are expected to be significantly smaller than
the expected limits by the future LHC program.

%======================
%======================

It is worth mentioning that the FCNC transitions can also be probed in $t \bar{t}$ production when a top quark decays
anomalously into  $q+\gamma$ or $q+Z$.
However, it has been found that the limits would be looser than the ones obtained in single top productions~\cite{AguilarSaavedra:2001ab}.
In case of signal observation, LHC would also be able to discriminate between anomalous $tuV$ and $tcV$  ($V = \gamma,Z$) using the charge ratio technique~\cite{Khatibi:2014via}.
As we already discussed, this would be possible at the FCC-ee by having an efficient of charm tagging technique.

%
%%%%%%%%%%%%%%%%%%%%%%%%%%%%%%%%%%%%%%%%%%    Summary and conclusions    %%%%%%%%%%%%%%%%%%%%%%%%%%%%%%%%%%%%%%%%%%%%%%%%%
%
\section{Summary and conclusions} \label{Conclusions}

Top quark flavor-changing neutral current interactions are extremely forbidden in the SM framework because of the GIM mechanism.
The SM predictions for branching ratios of the top quark decay into a photon or a $Z$ boson and an up-type
quark are at the order of $10^{-14}$. However, several extensions of the SM can enhance the
branching ratios by a factor of $10^{8-9}$ depending on the model. Therefore, precise measurement of
these branching ratios provide an excellent possibility to probe new physics beyond the SM in the top quark sector.
While it is impossible to measure the branching ratios with the precisions of order of $10^{-14}$ to test the SM,
observation of sizable branching ratios would indicate new physics beyond the SM.
FCC-ee with a clean environment and high luminosity would provide a unique opportunity to
measure the properties of top quark and its interactions. In this work, we have investigated the sensitivity
and discovery prospects of FCC-ee to the top quark FCNC transitions.
We have looked for the FCNC $tq \gamma$ and $tqZ$ couplings in single top-quark production in the 
process of $e^- + e^+ \rightarrow t \bar{q}+\bar{t}q$.
We perform the analysis in a model independent way using the effective Lagrangian approach
at the center-of-mass energies of $\sqrt{s}$ = 240, 350 and 500 GeV. In the analysis, we only consider the
leptonic (electron and muon) decay of the $W$ boson in the top quark decay. 
The {\sc delphes} package has been employed to account for the detector modeling.
The main background contribution is coming from  $W^\pm jj$  production
when the $W$ boson decays leptonically, i.e. $ e^+ e^- \rightarrow W^\pm jj \rightarrow  \ell^+ \nu_\ell j j  ( \ell^- \overline\nu_\ell j j )$. 
Other considered backgrounds in this analysis include the top quark pair events in semileptonic decay mode and
$Z \ell^\pm \ell^\pm$ (with hadronic decay of $Z$). 
A set of kinematic variables has been proposed as the input variables to a multivariate analysis for
discrimination of signal from background processes.
We find the $3 \sigma$ discovery ranges and the upper limits at 95\% CL for three signal scenarios versus the
integrated luminosity at the center-of-mass energies of 240, 350 and 500 GeV.
We find that with increasing the center-of-mass energy stronger bounds
would be reachable. With an integrated luminosity of 300 fb$^{-1}$ at the center-of-mass energy of 350 GeV,
upper limits of $3.43 \times 10^{-5}$,  $4.97 \times 10^{-5}$  would be obtained
on $Br(t \rightarrow q \gamma)$ and $Br(t \rightarrow q Z)$ ($\sigma_{\mu\nu}-$type), respectively.
A looser upper limit of $1.83 \times 10^{-4}$ on $Br(t \rightarrow q Z)$ with $\gamma_{\mu}-$type interaction
is obtained. It is found that a sensitivity of the order of $10^{-6}$ at high integrated luminosities would be achievable. 
The results of this study has been presented in the FCC-ee (TLEP) Physics
Workshop (TLEP9)~\cite{TLEP9}, FCC Week 2015~\cite{FCC-Week} and FCC-ee (TLEP) Physics meetings~\cite{meeting1,meeting2}.  We found that FCC-ee would be able to provide us stringent upper limits on the FCNC anomalous couplings and this work
could serve as a base for  more detailed studies in future in the FCC-ee project.

%
%%%%%%%%%%%%%%%%%%%%%%%%%%%%%%%%%%%%%%%%%%    Acknowledgments    %%%%%%%%%%%%%%%%%%%%%%%%%%%%%%%%%%%%%%%%%%%%%%%%%
%
\section*{Acknowledgments}

The authors are grateful to Patrizia Azzi and Freya Blekman and other FCC-ee colleagues for many useful discussions and comments.
Authors are thankful School of Particles and Accelerators, Institute for Research in Fundamental Sciences (IPM) for financially support of this project.
Hamzeh Khanpour also thanks the University of Science and Technology of Mazandaran for financial support provided for this research.

%
%%%%%%%%%%%%%%%%%%%%%%%%%%%%%%%%%%%%%%%%%%    Bibliography    %%%%%%%%%%%%%%%%%%%%%%%%%%%%%%%%%%%%%%%%%%%%%%%%%
%

\end{document}